\documentclass[sigconf]{acmart}
\usepackage{graphicx}
\usepackage[acronym]{glossaries}
\usepackage{enumitem}
\usepackage{url}

\makeglossaries

\copyrightyear{2026}
\acmYear{2026}
\setcopyright{cc}
\setcctype{by}
\acmConference[MODELS Companion 2026]{ACM/IEEE 29th International Conference on Model Driven Engineering Languages and Systems}{October 04--09, 2026}{Málaga, Spain}
\acmBooktitle{ACM/IEEE 29th International Conference on Model Driven Engineering Languages and Systems (MODELS Companion 2026), October 04--09, 2026, Málaga, Spain}
\acmDOI{10.1145/3837062.3839094}
\acmISBN{979-8-4007-2903-4/2026/10}

\setkeys{glslink}{hyper=false}
\newacronym{dt}{DT}{Digital Twin}
\newacronym{edt}{EDT}{Enterprise Digital Twin}
\newacronym{sots}{SoTS}{Systems of Twinned Systems}
\newacronym{pt}{PT}{Physical Twin}
\newacronym{dk}{DK}{Domain Knowledge}
\newacronym{ai}{AI}{Artificial Intelligence}
\newacronym{llm}{LLM}{Large Language Model}
\newacronym{kg}{KG}{Knowledge Graph}
\newacronym{rag}{RAG}{Retrieval-Augmented Generation}
\newacronym{mde}{MDE}{Model-Driven Engineering}
\newacronym{mbe}{MBE}{Model-Based Engineering}
\newacronym{oml}{OML}{Ontological Modeling Language}
\newacronym{owl}{OWL}{Web Ontology Language}
\newacronym{cps}{CPS}{Cyber-Physical System}
\newacronym{mbse}{MBSE}{Model-Based Software Engineering}
\newacronym{mbsye}{MBSyE}{Model-Based Systems Engineering}
\newacronym{dte}{DTE}{Digital Twin Engineering}
\newacronym{cot}{CoT}{Chain-of-Thought}
\newacronym{abs}{ABS}{Aspect-Based Summarization}
\newacronym{uml}{UML}{Unified Modeling Language}
\newacronym{cicd}{CI/CD}{Continuous Integration and Continuous Deployment}
\newacronym{dsl}{DSL}{Domain-Specific Language}
\newacronym{emf}{EMF}{Eclipse Modeling Framework}
\newacronym{nlp}{NLP}{Natural Language Processing}
\newacronym{ie}{IE}{Information Extraction}
\newacronym{ol}{OL}{Ontology Learning}
\newacronym{cq}{CQ}{Competency Question}
\newacronym{sar}{SAR}{Software Architecture Reconstruction}
\newacronym{dtdf}{DTDF}{Digital Twin Description Framework}
\newacronym{vnv}{V\&V}{Validation and Verification}
\newacronym{sye}{SyE}{Systems Engineering}
\newacronym{ra}{RA}{Reference Architecture}
\newacronym{sos}{SoS}{Systems-of-Systems}
\newacronym{ahase}{AHASE}{AI-Hyperagile Software Engineering}
\newacronym{bi}{BI}{Business Intelligence}
\newacronym{wia}{WIA}{What-If Analysis}
\newacronym{fm}{FM}{Foundation Model}
\newacronym{wm}{WM}{World Model}
\newacronym{tsfm}{TSFM}{Time-Series Foundation Model}
\newacronym{mvp}{MVP}{Minimum Viable Product}
\newacronym{mcmc}{MCMC}{Markov Chain Monte Carlo}
\newacronym{ui}{UI}{User Interface}
\newacronym{sme}{SME}{Subject Matter Expert}

\begin{document}

\title{Involving before Evolving: A Vision for Trustworthy Enterprise Digital Twin Engineering}

\author{K\'{e}rian Fiter}
\orcid{0009-0001-7731-0299}
\affiliation{
  \institution{Polytechnique Montr\'{e}al}
  \city{Montr\'{e}al}
  \country{Canada}
}
\email{kerian.fiter@polymtl.ca}

\author{Adil Lagrou}
\orcid{0009-0005-7955-0797}
\affiliation{
  \institution{Polytechnique Montr\'{e}al}
  \city{Montr\'{e}al}
  \country{Canada}
}
\email{adil.lagrou@polymtl.ca}

\author{Franck Dervault}
\orcid{0009-0003-8809-3490}
\affiliation{
  \institution{Michelin Canada}
  \city{Montr\'{e}al}
  \country{Canada}
}
\email{franck.dervault@michelin.com}

\author{Bentley Oakes}
\orcid{0000-0001-7558-1434}
\affiliation{
  \institution{Polytechnique Montr\'{e}al}
  \city{Montr\'{e}al}
  \country{Canada}
}
\email{bentley.oakes@polymtl.ca}

\date{October 2026}

\begin{abstract}
Enterprise Digital Twins (EDTs) promise data-driven decision support at organizational scale, but realizing them requires navigating siloed departments, tacit knowledge, and high-stakes decisions with long-horizon consequences. Existing approaches involve domain experts during model development but focus less on early organizational buy-in in EDTs. We present a vision for trustworthy EDT engineering grounded in an `involving before evolving' paradigm: rapidly involving stakeholders through a working prototype before evolving toward federation and full interoperability. Our three-stage approach combines foundation models for rapid prototyping, an ontological backbone for federated interoperability, and observability tooling for stakeholder trust. We ground our vision in an ongoing collaboration with Michelin, a multinational manufacturer, where an initial prototype has helped support stakeholder buy-in.
\end{abstract}

\begin{CCSXML}
<ccs2012>
   <concept>
       <concept_id>10002951.10003227.10003228</concept_id>
       <concept_desc>Information systems~Enterprise information systems</concept_desc>
       <concept_significance>500</concept_significance>
       </concept>
   <concept>
       <concept_id>10010147.10010178.10010187.10010195</concept_id>
       <concept_desc>Computing methodologies~Ontology engineering</concept_desc>
       <concept_significance>300</concept_significance>
       </concept>
 </ccs2012>
\end{CCSXML}

\ccsdesc[500]{Information systems~Enterprise information systems}
\ccsdesc[300]{Computing methodologies~Ontology engineering}
% \ccsdesc[300]{Computing methodologies~Modeling and simulation}

\keywords{Enterprise Digital Twin, What-if Analysis, Uncertainty, Trust, Federation, Ontologies, Foundation Models, Large Language Models}

\maketitle

\section{Introduction}

% \bentley{Notes from Maxime's presentation:
% Goal oriented req engineering with KAOS, resistance from stakeholders on rocking the boat, large organizations and municipalities are siloed, positive and negative consequences of services, aggregated/filtered data for privacy, stability over innovation, trustworthy DTs to be accepted, ranges rather than precise predictions (people don't trust precise predictions, they trust ranges and distributions more), moving from insights into manual actions (recommendations as a mid-way between insights and actions), mapping data between different systems to get started, people like maturity roadmaps, traceable and explainable decisions, sovereignty for data/processing, human in the loop rather than true DTs, capturing what data thé user looked at on that day, dyadic model, action design research}

A \gls{dt} is a virtual representation of a real system that enables its monitoring, prediction, and control through bidirectional data exchange~\cite{madni2019leveraging}.
\glspl{dt} can have vastly different scales, from simple twinned systems to complex \gls{sots}~\cite{adesanya2025systems}.
Here, we consider an \gls{edt} twinning an enterprise system, defined as ``a set of purposive analyzable and simulatable models representing the enterprise in order to mimic real-world phenomena,'' supporting ``what-if scenario[s] playing in virtual space''~\cite{kulkarni2019towards}.
Our \gls{edt} is a blend of an ``Acknowledged SoTS'' and ``Directed SoTS''~\cite{adesanya2025systems}, as there exist models, goals, and overall independence at both department- and enterprise-level.

While \glspl{dt} are defined by \textit{automatic} bidirectional data exchange \cite{kritzinger2018digital}, automatic control in \glspl{edt} is neither feasible nor desirable for high impact business decisions, implying the loop can only be closed at a long time-scale. We term this a \textit{long-loop} DT, where a human-in-the-loop affects the real system based on the DT, but at a longer time-scale (days, months, years) and with potentially irreversible consequences.
Thus, \glspl{edt} remain within the \gls{dt} paradigm: they receive enterprise data, use modeling, simulation, and AI techniques to predict behavior and prescribe actions, and close the loop through human-mediated organizational decisions. In particular, we focus on the lessons learned for the interoperable engineering of EDTs.

\paragraph{Context and Challenges}
We ground our vision in an ongoing collaboration with Michelin, a multinational manufacturer involved in world-wide supply chains, marketing, and sales, which serves as our motivating context.
Their business professionals operate across a complex structure with specialized \gls{dk} and responsibilities. They must navigate a siloed data and processes ecosystem to model business scenarios and assess cross-functional impacts. 
In particular, they rely on \textit{tacit knowledge} to perform what-if analysis~\cite{gathani2025if} and make decisions, which lacks explainability, data-based grounding, and is at risk of disappearing when individuals leave the organization.
This fragmented information environment limits the company’s ability to rapidly evaluate future scenarios, quantify uncertainty, and generate robust long-term forecasts for informed decision-making and organizational resilience.

We have extracted the following challenges for this EDT project through informal discussion with company stakeholders:
\begin{itemize}
    \item[C1] \textit{Decision-making support} to explore business strategies
    \item[C2] \textit{What-if simulations} of financial scenarios
    \item[C3] \textit{Easy modeling} by non-technical stakeholders
    \item[C4] \textit{Interoperability} across (systems-of-)systems
    \item[C5] \textit{Observability} of the solution and the scenarios
\end{itemize}

% The remainder of the paper presents our involving before evolving vision and its three components: rapid prototyping, ontology-backed federation, and observability for trustworthiness.
\begin{figure*}
    \centering
    \includegraphics[width=1\linewidth]{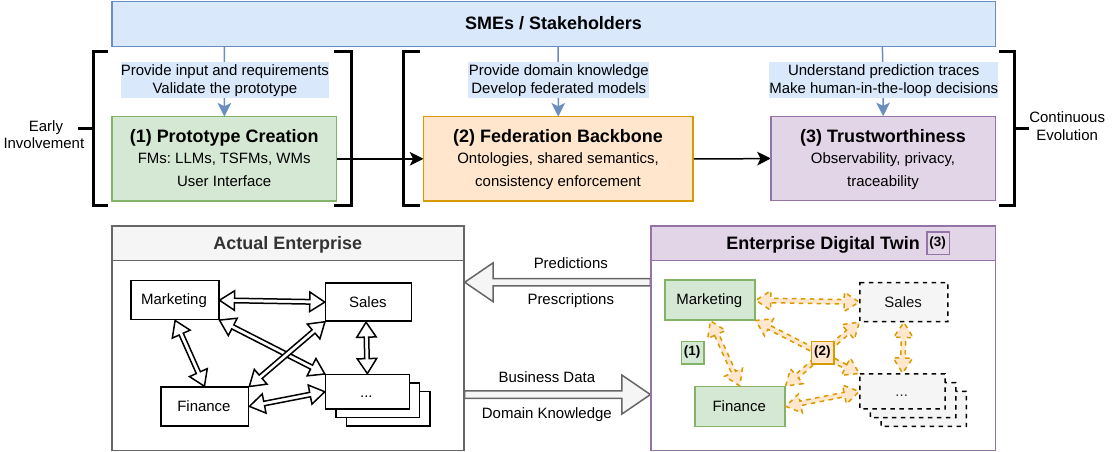}
    \caption{Outline for our \gls{edt} vision, reflecting our current progress and planned steps.}
    \label{fig:approach}
\end{figure*}

\section{Our Vision and Approach}\label{sec:approach}

Our vision is to develop a \textit{trustworthy} and \textit{interoperable} \gls{edt} for these departments to improve their prediction and decision-making capabilities. Large organizations can favor stability over innovation, and introducing a global \gls{edt} could face resistance from stakeholders wary of disrupting established workflows. We therefore argue for a socio-technical approach centered on an \textit{involving before evolving} paradigm to secure organizational buy-in, 
before evolving the \gls{edt} toward higher levels of maturity and automation~\cite{alskaif2025evolution}.
Our vision complements agent-based techniques~\cite{barat2019actor,barat2025constructing,barat2025synergic} by focusing on the earlier socio-technical problem of securing organizational buy-in.

This \textit{involving before evolving} vision is driven by the \textit{long-loop} nature of EDTs. In most DTs, there is a tight feedback loop between the \gls{pt} and DT~\cite{fitzgerald2024engineering}, where errors have immediate consequences. In a long-loop DT, a prediction could drive a decision with severe long-term consequences, such as restructuring a supply chain or opening a new market, which raises the bar for stakeholders to understand and trust the EDT and its insights.

\paragraph{Proposed Approach}

The three-stage approach is seen in Figure~\ref{fig:approach}:

\begin{enumerate}
    \item \textbf{Prototype Creation for Quick Involvement} (\textit{Targets C1, C2, C3}). Our vision is to quickly collect requirements from \glspl{sme}, choose a formalism, and build a prototype (a low-maturity early instance of the EDT) using \glspl{fm} to involve \glspl{sme} as soon as possible in the development process. We integrate \glspl{sme}' \gls{dk} into the simulation inputs and facilitate their validation of the low-maturity \gls{edt}.

    In our approach, we propose using \glspl{llm}, \glspl{tsfm}~\cite{kottapalli2025foundationmodelstimeseries}, and Monte Carlo simulation integrating uncertainty (ranges and distributions) to perform what-if financial simulations.
    
    \item \textbf{Federation for Interoperable Evolution} (\textit{Targets C4}).  
    Soon after prototype validation, our vision is to federate~\cite{berardinelli2026barriers} these models using an ontological backbone to formalize departmental \gls{dk}. These predictions should be based on models which are federated and owned by each department. This will enable a federated prototype with organizational buy-in (adoption and trust), even across departments with entrenched silos and differing agendas.
    
    Our approach proposes the \gls{oml}~\cite{elaasar2023opencaesar} and related methodologies~\cite{Oakes2024ServiceDriven,humphries2025integrated} to support federation, semantic alignment, and interoperability.
    
    \item \textbf{Observability for Promoting Trust} (\textit{Targets C5}). Finally, building on the federated models, we envision an ecosystem where data and predictions are explainable, traceable, versioned, access controlled, consistent, and observable across silos to promote trustworthiness in the resulting ecosystem.
    
    Our approach proposes observability metrics and development of analysis tools suited to \glspl{sme} (such as DTInsight~\cite{fiter2025dtinsight}) to increase the trustworthiness of the \gls{edt}.
\end{enumerate}

Currently, we have created a decision support prototype for the marketing and finance departments. Our next step is to build a federated ontological backbone, before modeling other departments and making the system trustworthy for decision support.

\section{Prototype Creation for Quick Involvement}\label{sec:rv1}

We propose to quickly capture requirements in an informal textual manner and build a minimum viable product to support early validation before investing in full verification and formalization~\cite{cederbladh2024early,cederbladh2025road}. While continuous \textit{evolving} of the DT remains key~\cite{alskaif2025evolution}, sustaining it requires buy-in from stakeholders in the form of a tangible prototype that immediately is \textit{involving} them and their knowledge.

To achieve rapid deployment, we propose to develop \glspl{ui} and models that would otherwise require expert \gls{dk} with \glspl{llm}, and use \glspl{fm} such as \glspl{tsfm}~\cite{kottapalli2025foundationmodelstimeseries} or action-conditioned \glspl{wm}~\cite{ding2025understanding}, as rapid forecasting or simulation placeholders where expert-developed models are not yet available.
\glspl{fm} are broadly trained, large-scale models that can be adapted to many downstream problems: \glspl{llm} are \glspl{fm} specialized for language understanding and generation; \glspl{tsfm} are \glspl{fm} specialized for temporal or sequential data using transformer architectures; and action-conditioned \glspl{wm} are \glspl{fm} specialized for learning compact predictive representations of environments and forecasting future states given current observations and planned actions.
In particular, we go further than Giabbanelli \textit{et al.}, who proposed \glspl{llm} as tool translators between models and simulation tools~\cite{giabbanelli2025over}, and directly build models, tools, and interfaces with various \glspl{fm}.

The outcome of this approach is to lift user interactions from error-prone and inconsistent spreadsheet calculations to automated and visual simulations to prioritize business-level decisions.

\begin{figure}[htbp]
    \centering
    \includegraphics[width=1\linewidth]{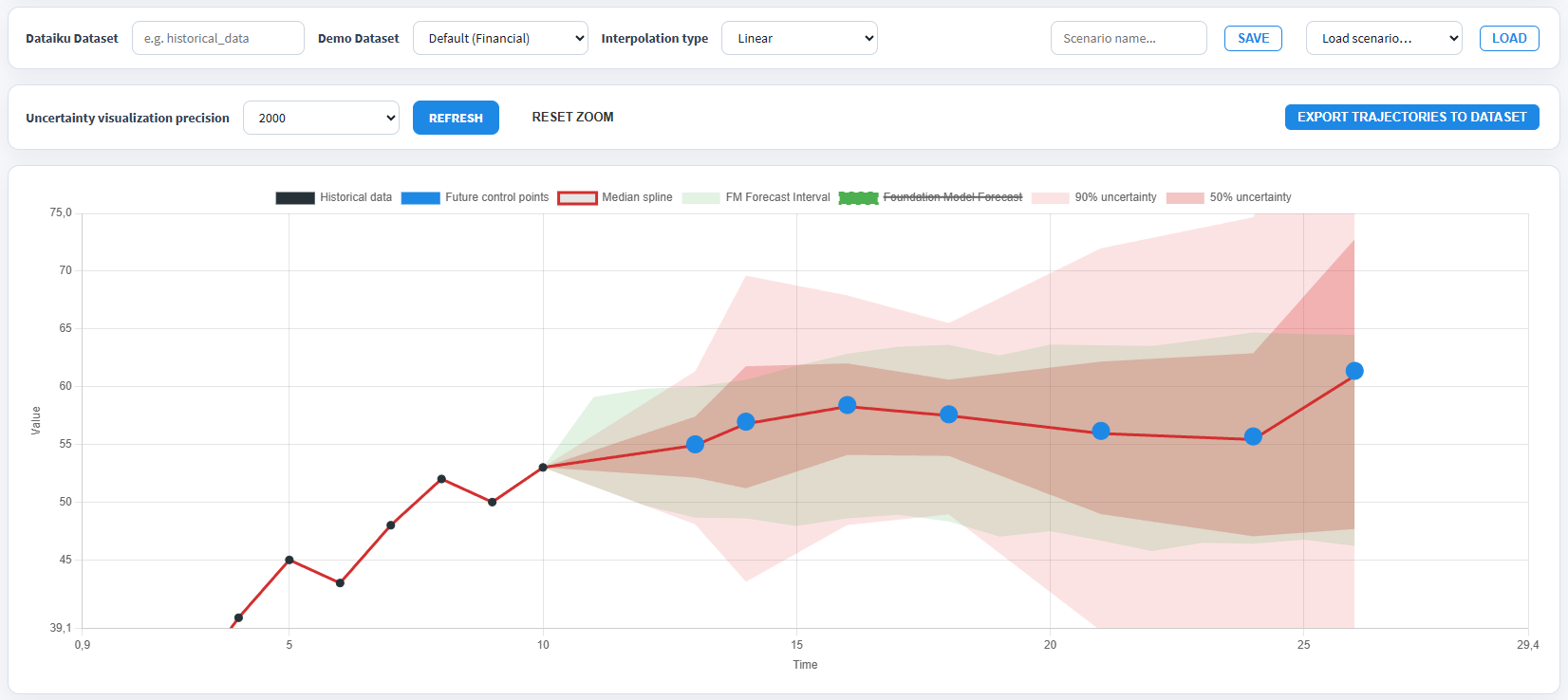}
    \caption{UI prototype for integrating historical data,  \glspl{tsfm}, and uncertain trajectory predictions}
    \label{fig:ui}
\end{figure}

\paragraph{Motivating Example}
We use our industrial context to illustrate this stage: after capturing informal requirements from multiple departments, we have built a light-weight tool whose \gls{ui} is showcased in Figure~\ref{fig:ui}. It takes historical time-series data as input and proposes a spline trajectory to the user based on the predictions of a locally-hosted \gls{tsfm}~\cite{das2023decoder} (\texttt{TimesFM 2.5}\footnote{\url{https://github.com/google-research/timesfm/}}, 10/90th percentiles in green). The user can manually edit the spline points (in blue) and select the uncertainty range (10/90th percentiles in light red, 50th percentile in red) to create their trajectory prediction.
This provides users a visual and real-time interface to easily embed their \gls{dk} predictions.
We then use Monte Carlo simulation to sample from these spline-defined uncertainty ranges and propagate uncertainty through downstream expert-developed models, producing distributions over predicted business outcomes.

This functional prototype,  rapidly built using agentic LLMs hosted on company infrastructure, satisfies \gls{sme} requirements and demonstrates a novel way to easily model scenarios from historical data with uncertainty control.
This prototype focuses on clear visual understanding to support financial and marketing decision-making over the short and long term.

Despite this simple formalism, it has been effective in securing buy-in and motivating further discussion across marketing and finance departments.
It led to the refinement and creation of requirements, such as how the tool could represent multiple datasets to detect correlations, and how the \gls{tsfm} could be fine-tuned or improved to deliver more precise predictions.
This prototype will be extended to the ontological backbone described in the next section.
 \section{Federation for Interoperable Evolution}\label{sec:rv2}

%\bentley{Define interoperability: \url{https://www.tandfonline.com/doi/epdf/10.1080/0951192X.2020.1736636?needAccess=true}}

% \bentley{To read: \url{https://www.researchgate.net/publication/407296265_Driving_Optimization_Through_an_Ontological_Digital_Engineering_Workflow}}
% \kerian{Talks about technical (infra level) and data interoperability} Presentation: \url{https://ncor-network.org/presentations/stids2026/Track%20B%20Talk%3A%20Ontology-Driven%20Test%20Strategy%20Optimization.pdf}

% \todo{value in taxonomy creation so that one department can understand what others have created}

% \todo{cite \cite{barat2020orgml} OrgML}

% \todo{decide on how to use ontologies for federation and semantic interoperability}

Interoperability is ``the ability of systems to exchange information and use the information that has been exchanged''~\cite{Leal03032020}. To realize this across departmental silos, we adopt a federated approach that reflects the organization's social structure, following ``business and architecture isomorphism''~\cite{Thones2015microservices}, as demonstrated in organizational modeling approaches such as OrgML~\cite{barat2020orgml}. 
Rather than a federated DT in the sense of Navarro \textit{et al.} (``a composition of autonomous DTs''~\cite{navarro2026towards}), we pursue a DT characterized by coordinated autonomy. 

We aim for federation to let teams retain ownership of their data, models and services, while committing to shared semantic contracts that guarantee interoperability. Here, ontologies act as the backbone of these contracts in three main roles: (1)~\textit{structural wiring} of components and their relations, ensuring that when departments connect their DTs, the ontology defines how each component can legally be composed, preventing structurally invalid federations from being created in the first place; (2)~\textit{typed interfaces} so that data exchanged between departments carries an ontologically grounded type to reduce ambiguity~\cite{gregory2026opt} (e.g., when marketing sends a demand signal to the supply chain, both sides are aligned on what that signal represents); (3)~\textit{domain semantics} expressed as constraints and rules to enable automated consistency checks and inference 
% \adil{using retyping/reclassification} 
(e.g., the supply chain may only act on a demand forecast if it has been approved by finance, enabling automated reasoning to catch any violation of this sort without human auditing). 

% \kerian{could it help differentiate "free demand" from the markets with "constrained demand" once logistics and other are taken in account (Supply chain)}

% Structural wiring ensures that when departments connect their DTs, the ontology defines which components can legally be composed and how, preventing structurally invalid federations from being created in the first place. Typed Interfaces mean that every message exchanged between department DTs carries an ontologically grounded type, so for example when marketing sends a demand signal to Supply chain, both sides are aligned on what that signal represents. And finally, Domain semantics go even further, by encoding enterprise-level rules and constraints. for example, that Supply chain may only act on a demand forecast if it comes from an approved Finance source, thus enabling automated reasoning to catch any violation of this sort without human auditing.

% \bentley{I would expect these three roles to be the next paragraph}. 

Beyond federated DTs~\cite{liu2024review}, our Federated \gls{edt} approach mirrors the enterprise’s organization that is split into departmental silos with distinct agendas, sets of tools, and \gls{dk}. These departments operate over large-scale datasets that are intentionally segregated to preserve confidentiality and ensure accountability. As the silos evolve independently over time, it is essential to specify and maintain stable versioning and well-defined interfaces across their models and systems to ensure interoperability and reliability for cross-cutting, multi-department what-if analysis.

Our approach, based on the proposal from Humphries \textit{et al.}~\cite{humphries2025integrated}, suggests that \glspl{sme} from each department express their \gls{dk} as an owned  ontology, interoperable with the others through shared contracts and rules. The key insight is in how we use the ontologies: using them merely as \textit{taxonomies} would resolve terminology clashes between departments but would not support reasoning. This is performed by encoding domain semantics as inference rules within each department's own vocabulary.

For example, historical data can require different fidelity depending on the domain in which it is applied. One market may exhibit long cycles and therefore require ten years of historical data, whereas another only needs six months. Capturing this \gls{dk} in ontologies lets us reason over models that use multiple historical inputs and determine whether the overall fidelity is adequate. If a model draws on two high-fidelity inputs and one low-fidelity input, the reasoning layer could flag the model's output as degraded.

Crucially, vocabulary owners must be able to author and understand their own vocabularies. We therefore adopt  \gls{oml}, a \gls{dsl} over OWL for defining concepts, relationships, constraints, and basic rules, as well as associated tooling~\cite{elaasar2023opencaesar}. This reduces the friction required to develop ontologies in OWL or Prot\'eg\'e~\cite{protege}, yet preserves the semantic rigor needed for typed interfaces and automated checks. To further aid development, we can layer a \gls{dsl} or generation tool on top of OML as needed.
\section{Observability for Promoting Trust}\label{sec:rv3}

\glspl{sme} rely on their experience to make business decisions. Therefore, they are wary of decision-support tooling unless they trust its predictions. We define trustworthiness as the ability of stakeholders to inspect, understand, trace, and challenge \gls{edt} outputs, while preserving data sovereignty and organizational accountability.

In our industrial example, we support trustworthiness through \textit{observability}~\cite{majors2022observability} by employing an explainable formalism based on probability (Monte Carlo simulation) that can be analyzed through techniques such as sensitivity analysis. We keep the stakeholders in control by allowing them to define inputs through curve editing (see Figure~\ref{fig:ui}), which aligns with their \gls{dk} understanding, i.e., market tendencies and seasonal trends. Furthermore, our prediction module will feature versioning, ensuring stakeholders understand which tool version and datasets were used to generate the final prediction.

Beyond observability and human oversight, \gls{edt} trustworthiness also depends on the privacy and safety of the supporting tooling, since the \gls{edt} processes critical business data. This requires local \glspl{fm} or models that ensure the organization’s data sovereignty. We employ agentic LLMs hosted on the company’s infrastructure to develop the tool, and a local \gls{tsfm} instance runs in a container to generate time-series forecasts on historical and sensitive data.

To broaden this trust to the entire \gls{edt} development life-cycle, we plan to enhance our DT observability tool DTInsight~\cite{fiter2025dtinsight} with lightweight \gls{oml} modeling capabilities. This aligns with our vision of continuously generating a ``constellation'' (a conceptual visualization of a \gls{dt})~\cite{Gil2024towardssystematicreporting} alongside a dynamic reporting page that tracks the \gls{edt}'s evolution, thereby enabling stakeholders to actively collaborate on and discuss the system as it is being built.
\section{Discussion} \label{sec:conclusion}

\paragraph{Expected benefits}

The expected benefits of our vision are (1)~rapid and increased buy-in in \glspl{edt} from stakeholders through early involvement with a prototype; (2)~reduced errors and better collaboration between departments through ontology-backed semantic alignment; (3)~increased agility due to the federated approach that allows departments to focus on and control their data and models; (4)~trusted probabilistic what-if simulations that are observable, traceable, and privacy-respecting; (5)~increased overall resilience for the enterprise due to data-driven decision-making support.

\textit{Research agenda alignment.} Our model-based \gls{dt} engineering survey~\cite{Zech2026MBDTE} found an open research challenge in making Model-Based Engineering tooling accessible to non-experts, and we extend that concern to organizational adoption. Our vision further aligns with Combemale \textit{et al.}'s \gls{dt} engineering roadmap~\cite{combemale2025engineering}:
\begin{itemize}[leftmargin=1.3cm]
    \item[\textbf{RQ\_P2}] ``What communication tools and practices can enhance collaboration between data scientists, engineers, and operations teams in a DT engineering environment?''
    \item[\textbf{RQ\_P3}] ``What communication strategies can ensure that all stakeholders are informed and engaged throughout the DT models lifecycle? [...]''
\end{itemize}

\paragraph{Related approaches}

Our work relates to Franzén's exploration of design and trade spaces in early \gls{sos} development~\cite{franzen2023system}. They capture needs in an ontology that outputs a reduced design space through reasoning. \gls{sos} simulations generate performance measures to identify the best architecture, which can then be translated into interactive visual analytics. Conversely, we focus on \glspl{edt} and propose to involve stakeholders early through a prototype that supports iterative design-space exploration and organizational adoption, using ontologies to federate the \gls{sos}.

Barat \textit{et al.} have extensive work on \glspl{edt}~\cite{barat2019actor,barat2022enterprise,barat2025synergic}. However, they focus on agent-based techniques using the Enterprise Simulation Language (ESL)~\cite{clark2017esl}. Instead, we propose a top-down approach that considers the \gls{edt} as a digital wrapper over the social structure of the organization with simulation placeholders. We provide the formalism (Monte Carlo simulation) and interfaces for \glspl{sme} to work within their \gls{dk} in a federated manner. Barat \textit{et al.} also touch on the need for \glspl{sme} to increasingly focus on validation due to improving LLM modeling capabilities~\cite{barat2025constructing}. Our vision goes further into prototyping and trustworthiness concerns.

We also see potential in action-conditioned \glspl{wm} as placeholders for simulating business behavior and finding optimal scenarios. These models are currently developed mostly on video data, but we expect their generalized ability to explore possible futures will be useful for businesses~\cite{hafner2025mastering}.

\section{Conclusion}

This paper has explored our vision of \textit{involving before evolving}, where we target non-technical stakeholders divided among organizational silos.
Currently, we have a working \gls{edt} prototype with stakeholder buy-in across Michelin departments.

The next step is to connect this prototype to the ontological backbone, using the methodological framework of Humphries \textit{et al.}~\cite{humphries2025integrated}. This will support versioned releases of models and datasets that can be composed into a trustworthy \gls{edt}.
We also plan to make requirements gathering more incremental by progressively formalizing requirements against the emerging ontology.
Although this reverses the order adopted in some ontology-first approaches~\cite{franzen2023system}, it enables discussions to start from a validated prototype and gradually converge toward a shared formal language.

\textit{Open challenges.} First, it is unclear whether ontology-backed contracts can handle real semantic heterogeneity at scale, or whether automated consistency checks remain feasible as the federation grows. Second, scenario provenance, i.e., labeling, sharing, and comparing trajectory scenarios across departments over time, is important for EDTs but is not yet handled in our approach. Third, we assume that EDT-informed decisions will produce better organizational outcomes. However, the long-loop time-scale makes this methodologically difficult to evaluate without a longitudinal study across multiple deployments, which is not feasible in our project.

\begin{acks}

Feedback on the \textit{long-loop DT} concept came from discussions with Istvan David (McMaster University), Houari Sahraoui (Universit\'{e} de Montr\'{e}al), and Maxime Lamothe (Polytechnique Montr\'{e}al). We also sincerely thank our partner Michelin, Mitacs Accelerate (IT38322), and NSERC Discovery (RGPIN-2024-05622) for funding support.

\end{acks}

\bibliographystyle{ACM-Reference-Format}
\bibliography{main}

\end{document}